# Two-dimensional honeycomb borophene oxide: Strong anisotropy and nodal loop transformation

Chengyong Zhong[1,2,*], Weikang Wu[3], Junjie He[1,2], Guangqian Ding[2], Yi Liu[2], Dengfeng Li[2]

Shengyuan A. Yang[3,4,*], and Gang Zhang[5,*]

[1] *Institute for Advanced Study, Chengdu University, Chengdu 610106, China*

[2]*School of Science, Chongqing University of Posts and Telecommunications, Chongqing 400065, China*

[3] *Research Laboratory for Quantum Materials, Singapore University of Technology and Design, Singapore 487372, Singapore*

[4] *Center for Quantum Transport and Thermal Energy Science, School of Physics and Technology, Nanjing Normal University, Nanjing 210023, China*

[5] *Institute of High-Performance Computing, A*STAR, 138632, Singapore*

---

[*] Corresponding Author:

zhongcy90@gmail.com;  shengyuan_yang@sutd.edu.sg; zhangg@ihpc.a-star.edu.sg

# Abstract

The search for topological semimetals is mainly focused on heavy-element compounds as following the footsteps of previous research on topological insulators, with less attention on light-element materials. However, the negligible spin orbit coupling with light elements may turn out to be beneficial for realizing topological band features. Here, using first-principles calculations, we propose a new two-dimensional light-element material—the honeycomb borophene oxide (h-$B_2O$), which has nontrivial topological properties. The proposed structure is based on the recently synthesized honeycomb borophene on Al (111) substrate [*Sci. Bull.* 63, 282 (2018)]. The h-$B_2O$ monolayer is completely flat, unlike the oxides of graphene or silicene. We systematically investigate the structural properties of h-$B_2O$, and find that it has very good stability and exhibits significant mechanical anisotropy. Interestingly, the electronic band structure of h-$B_2O$ hosts a nodal loop centered around the Y point in the Brillouin zone, protected by the mirror symmetry. Furthermore, under moderate lattice strain, the single nodal loop can be transformed into two loops, each penetrating through the Brillouin zone. The loops before and after the transition are characterized by different $\mathbb{Z} \times \mathbb{Z}$ topological indices. Our work not only predicts a new two-dimensional material with interesting physical properties, but also offers an alternative approach to search for new topological phases in 2D light-element systems.

# Introduction

Over past decades, one of most profound leaps in condensed matter physics is the recognition of non-trivial band structures in solid materials, leading to a field of topological materials[1-5]. At present, the study on topological materials is mainly focused on compounds involving heavy elements, hoping that the strong spin-orbit coupling (SOC) could help to induce the desired topological band features[6-10]. However, the nontrivial topology can also appear in light-element materials with negligible SOC. For example, the band inversion and associated topological feature for graphene have been noticed long ago[11,12]. Recent works also revealed a variety of topological nodal-point[13], nodal-loop (including Hopf-link)[14-17], or even nodal-surface[18,19] semimetal phases in carbon allotrope materials. It has been emphasized that such topological phases in the absence of SOC is fundamentally distinct from the SOC-induced topological phases[17]. Without SOC, the real spin is a dumb degree of freedom[20]. Then the fundamental time reversal operation ($\mathcal{T}$) satisfies $\mathcal{T}^2 = 1$, contrasting with $\mathcal{T}^2 = -1$ for the spinful case. Thus, the two belong to different topological classifications[21].

Boron is next to carbon in the periodic table and is also a typical light element. Previous works have shown that boron can also form various allotropic structures, and some of them also exhibit topological properties. For example, Dirac-type band crossings have been predicted in 2D $\beta_{12}$ borophene[22,23]. On the experimental side, the successful realization of 2D borophene[24,25] has triggered a lot of interest in searching for topological properties in 2D boron allotropes as well as their derivatives[26-34].

Considering the electron-deficiency in boron as compared with carbon, it seems that a honeycomb structure similar to graphene cannot be realized for 2D boron[35]. Surprisingly, a recent experimental work by Li et.al[36] reported the growth of such honeycomb borophene on Al(111) substrate. In view of the topological features in graphene, one may naturally ask: Does the honeycomb borophene also possess nontrivial topological features near the Fermi level? Unfortunately, for pristine honeycomb borophene, the answer is negative[35, 37].

Generally speaking, 2D materials are prone to oxidation under ambient conditions. For many cases, this is considered as a disadvantage, but from another point of view, oxidation can be used to improve the stability of the structure, and to tailor the physical and chemical properties. Moreover, oxygen is also a light element which can be helpful for achieving SOC-free topological materials. For instance, oxidized blue phosphene was found to host 2D double Weyl and psedospin-1 fermions[38]. The semi-Dirac semimetal was also reported in silicene oxide[39]. From the discussion, it is natural to consider oxidized borophene structures, which may have improved stability and possible topological band features.

Motivated by the recent experimental progress with the honeycomb borophene, here, through first-principle calculations, we propose a new 2D material: the honeycomb borophene oxide (h-$B_2O$). We find that in the fully optimized structure, h-$B_2O$ takes a strictly planar form, unlike the graphene or silicene oxides. The stability of h-$B_2O$ has been thoroughly checked via lattice dynamics and ab initio molecular dynamics calculations. At equilibrium state, the material possesses prominent anisotropic mechanical properties. Interestingly, h-B$_2$O hosts a single nodal loop centered around the Y point near the Fermi level. The nodal loop is protected by mirror symmetry, and we construct a simple tight-binding (TB) model to describe the low-energy band structure. Moreover, we find that under moderate strain, the single loop can be transformed into two loops, each penetrating through the Brillouin zone. The loops before and after the transition are characterized by different $\mathbb{Z} \times \mathbb{Z}$ topological indices. Our work reveals a new 2D topological material, which possess excellent mechanical and electronic properties and may serve as a promising platform for studying 2D topological phases.

## Results and Discussion

It is well known that oxygen atoms tend to adopt the bridge sites in graphene or silicene oxides, forming an epoxide structure[40-43]. As for the honeycomb borophene structure, our calculation shows that after adding an oxygen atom to the bridge site, the structure automatically relaxes, and the oxygen atom moves into the basal plane to form a completely planar structure (see Fig S1 of supporting information in detail). After optimization, we find one stable 2D borophene oxide structure, as shown in Figure 1(a). One observes that the monolayer is completely planar, and consists of hexagons elongated along the *b*-axis. Hence, we name this material as the hexagonal borophene oxide (h-B$_2$O).

The material has the space group *Cmmm* (No. 65), with two B and one O atoms in a primitive cell (see Fig 1a). The optimized lattice constants are a = 2.806 Å and b=7.127 Å, respectively. The equilibrium bond lengths of B-B and B-O are 1.706 Å and 1.341 Å, respectively. The corresponding Brillouin zone (BZ) is a deformed hexagon (see Fig 1b).

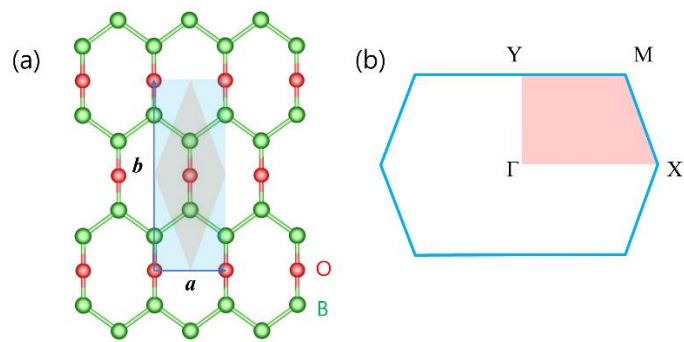

Figure 1. (a) The atomic structure of h-$B_2O$, the blue (rectangular) and orange (rhombus) shadowed areas stand for the conventional cell and the primitive cell, respectively. (b) The first BZ of h-$B_2O$ with the irreducible BZ colored in red.

We check the dynamic stability of h-$B_2O$ by calculating its phonon spectrum. The result is shown in Figure 2a. One can observe that there is no soft mode throughout the entire BZ, demonstrating the h-$B_2O$ is dynamically stable. In the vicinity of the Γ point, frequencies of out-of-plane acoustic phonons, known as the ZA phonons, show a parabolic dependence on the wave vector, which is a characteristic feature for layered materials[38, 44]. The linear dispersions are observed for the other two branches of in-plane acoustic phonons. The speed of sound (i.e. the slope of the acoustic branches at Γ point) is quite different along Γ-X (**a**-direction) and Γ-Y (**b**-direction), indicating a strong in-plane anisotropy, as we will discuss later. The thermodynamic stability of h-$B_2O$ is also verified through the *ab-initio* molecular dynamics (AIMD) simulations at 500 K (see Fig 2b, c) in the framework of NVT ensemble. We find that the structure well maintained after 10 ps, with a time step of 1 fs. Higher temperatures up to 1000 K are also tested, showing that the structure is still well kept (see Fig S2).

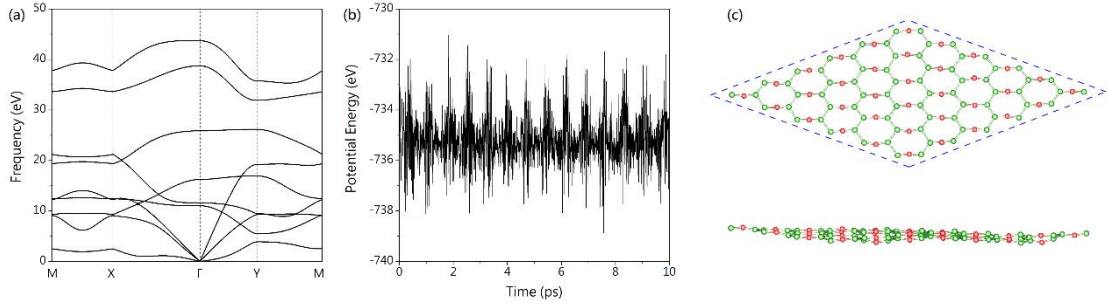

Figure 2. (a) The phonon spectrum of h-B$_2$O. (b) The total potential energy fluctuation of h-B$_2$O during AIMD simulation at 500 K. (c) The atomic configuration of h-B$_2$O after 10 ps of AIMD simulation at 500 K.

We also notice that a honeycomb borophene hydride (h-B$_2$H$_2$) with two hydrogen atoms located at both sides of the bridge sites of honeycomb borophene has been synthesized recently[45]. This h-B$_2$H$_2$ has identical space group with h-B$_2$O (see Fig S3). We have calculated the formation energies of these two structures with the formula: $\Delta E = [E_{Bx} - (\mu_B * n_B + \mu_x * n_x)]/(n_B + n_x), x \in \{H, O\}$ in, $\mu_B$ and $\mu_x$ are the chemical potential of B atom and X atoms, which are calculated from the hexagonal borophene and X$_2$ molecule respectively, $E_{Bx}$ is the total energy of system, $n_B$ and $n_x$ are the number of B atoms and X atoms in the structure. With this definition, a smaller value indicates a higher stability. The calculated formation energy for h-B$_2$O is -1.545 eV/atom, which is more than three times larger than that for h-B$_2$H$_2$ (-0.462 eV/atom). This implies that h-B$_2$O also has a good possibility to be realized in experiment. Furthermore, compared with the formation energies of other 2D B$_n$O$_m$ materials[46] as shown in Figure S4, it is clear that the predicted h-B$_2$O is located on the convex hull, indicating its stability.

Due to the lack of one electron compared with carbon, unlike graphene, the 2D honeycomb boron structure is typically unstable. As pointed out by Zhang *et al.*[47] and Shirodkar *et al.*[48], a possible approach to stabilize honeycomb boron is by electron doping. The recent experimental breakthrough by Li *et al*[36]. achieved this by growing honeycomb boron on the Al substrate. In this work, we offered another approach, i.e. by oxidation, which also makes a stable 2D planar structure.

Next, we examine the elastic properties. The elastic constants for h-$B_2O$ have been obtained from strain-energy-vs-strain curves, corresponding to suitable sets of deformations applied to a single unit cell[49]. The conventional rectangular lattice requires the independent elastic constants and strain energy to satisfy the relation: $E_s = \frac{1}{2}C_{11}\varepsilon_{aa}^2 + \frac{1}{2}C_{22}\varepsilon_{bb}^2 + C_{12}\varepsilon_{aa}\varepsilon_{bb} + 2C_{44}\varepsilon_{ab}^2$, where $E_s$ is the strain energy (the total energy of the strained state minus that of the equilibrium state) per unit area, $\varepsilon_{aa}$ and $\varepsilon_{bb}$ are the small strains along **a** and **b** directions in the harmonic regime, $\varepsilon_{ab}$ is the shear strain. $C_{11}$, $C_{22}$, $C_{12}$ and $C_{44}$ are the independent elastic constants, corresponding to the second partial derivative of strain energy with respect to the applied strain. Through fitting the strain energy curves associated with uniaxial, biaxial and shear strains (see Fig. 3a), the elastic constants are obtained: $C_{11} = 44.13$ N/m, $C_{22} = 237.41$ N/m, $C_{12} = 65.11$ N/m and $C_{44} = 11.45$ N/m. Clearly, those results satisfy the Born-Huang criteria[50] for rectangular cell: $C_{11}C_{22} - C_{12}^2 > 0$ and $C_{44} > 0$, indicating that h-$B_2O$ is also mechanically stable. We note that the calculated elastic constants $C_{11}$ and $C_{22}$ of h-$B_2O$ are larger and the mechanical anisotropy is more prominent as compared to phosphorene, which has been demonstrated to have remarkable anisotropic mechanical properties ($C_{11}$=24 N/m, $C_{22}$=103 N/m)[51]. Here, for comparison, we also repeat the calculation for h-$B_2H_2$, the result is shown in Fig 3b and Table 1.

Under orthogonal lattice, with the elastic constants, the in-plane Young's modulus Y and Possion ratio $\nu$ along an arbitrary direction specified with polar angle $\theta$ (here, $\theta$ is the angle relative to the **a** direction) can be calculated as follows[49]:

$$Y(\theta) = \frac{\Delta}{C_{11}s^4 + C_{22}c^4 + \left(\frac{\Delta}{C_{44}} - 2C_{12}\right)c^2s^2} \qquad (1)$$

$$\nu(\theta) = \frac{\left(C_{11} + C_{22} - \frac{\Delta}{C_{44}}\right)c^2s^2 - C_{12}(c^4 + s^4)}{C_{11}s^4 + C_{22}c^4 + \left(\frac{\Delta}{C_{44}} - 2C_{12}\right)c^2s^2} \qquad (2)$$

where $\Delta = C_{11}C_{22} - C_{12}^2$, $c = \cos(\theta)$ and $s = \sin(\theta)$. Through Eqs. (1) and (2) and using the elastic constants reported in Table 1, the $\theta$ dependence of Y and $\nu$ for h-$B_2O$ and h-$B_2H_2$ are plotted with polar coordinates in Figure 4. In such a plot, a fully isotropic elastic behavior is represented by a perfectly circular shape for Y and $\nu$. In contrast, the shapes of the Y and $\nu$ in Fig. 4 are highly anisotropic, with remarkable major axis and minor axis difference for h-$B_2O$, indicating that the mechanical properties of h-$B_2O$ exhibit extremely strong anisotropy. The anisotropy for h-$B_2O$ is stronger than that of the recent realized h-$B_2H_2$ (see Fig 4). The remarkable anisotropic elastic behavior for h-$B_2O$ can be partly attributed to the stiffening effect of B-O bonds that are exactly oriented along the **b**-direction (see Fig 1a). The B-O bond length (1.341 Å) is much shorter than the B-B bond (1.706 Å) oriented along the **a**-direction, suggesting that the bond strength of B-O is stronger than that of the B-B bond.

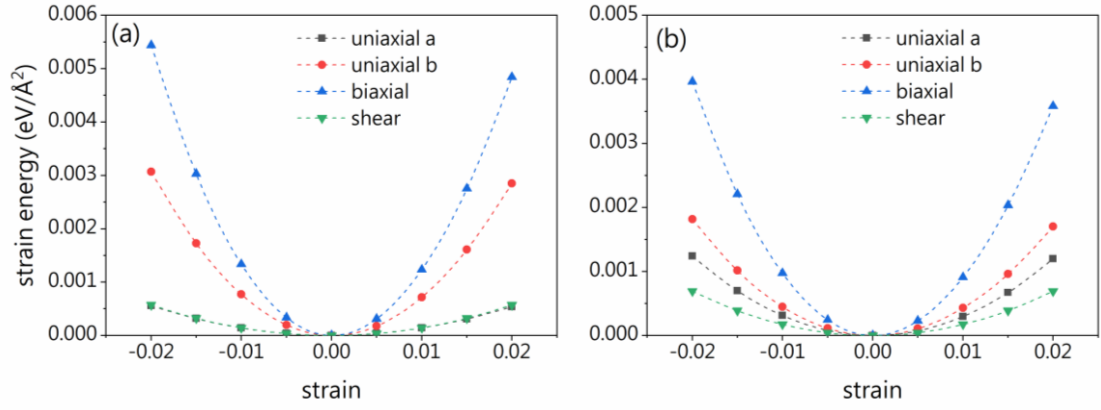

Figure 3. The strain energy per area under the uniaxial, biaxial and shear strains for (a) h-$B_2O$ and (b) h-$B_2H_2$.

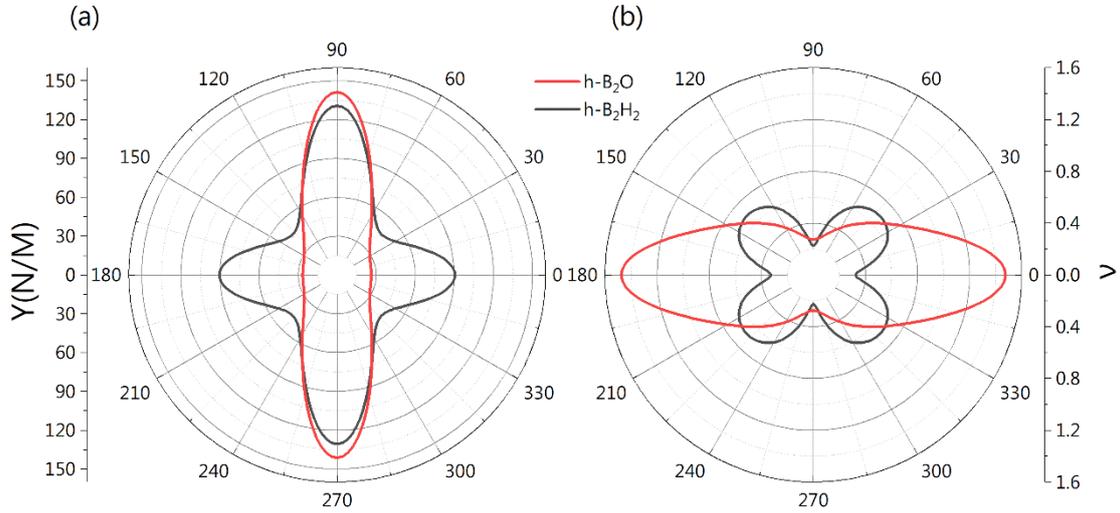

Figure 4. The $\theta$ dependence of (a) Y and (b) ν for h-$B_2O$ and h-$B_2H_2$.

Table 1. Space group (SG), elastic constants (N/m), Young's modulus (N/m), Possion ratio along **a/b** direction, and formation energy (eV/atom) for the two materials h-$B_2O$ and h-$B_2H_2$.

| System | SG | $C_{11}$ | $C_{22}$ | $C_{12}$ | $C_{44}$ | $Y_a$ | $Y_b$ | $\nu_a$ | $\nu_b$ | $\Delta E$ |
|---|---|---|---|---|---|---|---|---|---|---|
| h-$B_2O$ | Cmmm | 44.13 | 237.41 | 65.11 | 11.45 | 26.27 | 141.35 | 0.27 | 1.48 | 1.545 |
| h-$B_2H_2$ | Cmmm | 97.63 | 140.99 | 31.77 | 13.83 | 90.47 | 130.65 | 0.23 | 0.33 | 0.462 |

Now we investigate the electronic properties for h-$B_2O$. Its band structure calculated from first-principles calculations is plotted in Figure 5. One observes that two bands cross near the Fermi level. The orbital-resolved band structure manifest clearly that the two crossing bands have different orbital origins: One band is mainly contributed by the $p_y$ orbitals of B and O, while the other band is mainly from the $p_z$ orbitals of B and O (see Fig. 5a). The orbital constitution is further consolidated by the projected density of states (PDOS, see Fig. 5a) and the band decomposed charge densities (see Fig. 5b). A careful scan of the band structure shows that the crossing points on Γ-Y and Y-M are not isolated. They actually belong to the same nodal loop centered around the Y point (see Fig. 5c). It shows that the band ordering for the two low-energy bands are inverted between Y point and other high-symmetry points of the BZ, with the band inversion gap reaching 4.78 eV at the Y point. We also calculate the band structure by the more accurate HSE calculation (see Fig. S5), which confirms that the key band features are maintained. Therefore, we will focus on the PBE result in the following.

This nodal loop is protected by the mirror symmetry $M_z$. The $p_y$ orbital is even with respect to mirror operation $M_z$, however, the $p_z$ orbital is odd. Thus, the two crossing bands would have opposite mirror eigenvalues, such that they cannot hybridize and the crossing between them cannot be gapped. This has been checked from our first-principles calculations. Note that the protection holds when the SOC is negligible, which is the case for h-$B_2O$ since only light elements are involved (see Fig. S6). In the case of strong SOC, a gap can be opened at the nodal loop.

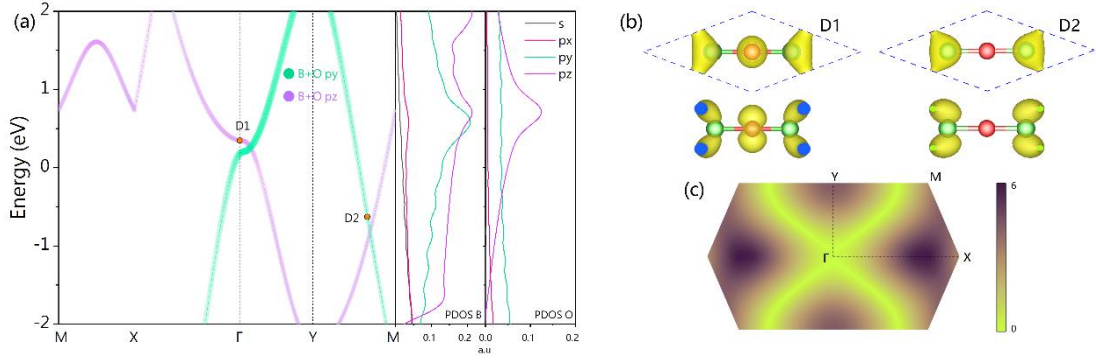

Figure 5. The electronic properties of h-B$_2$O. (a) The orbital-resolved band structure and PDOS. The green and purple dots represent the contributions of p$_y$ and p$_z$ orbital, respectively. (b) The band decomposed charge densities corresponding to two states as marked in (a). (c) The energy difference of p$_y$-band and p$_z$-band in the 2D BZ, the glowing yellow line indicates the location of the nodal loop.

Depending on the type of dispersion, nodal loops can be classified as type-I, type-II[52], or hybrid type[53]. Typically, a type-I loop is formed by the crossing between an electron-like band and a hole-like band; a type-II loop is formed by the crossing between two electron-like bands or two hole-like bands; and a hybrid loop can appear when one of the crossing bands has a saddle-type dispersion. Here, by examining the dispersion around the nodal loop, we find that the loop in h-B$_2$O belongs to type-I.

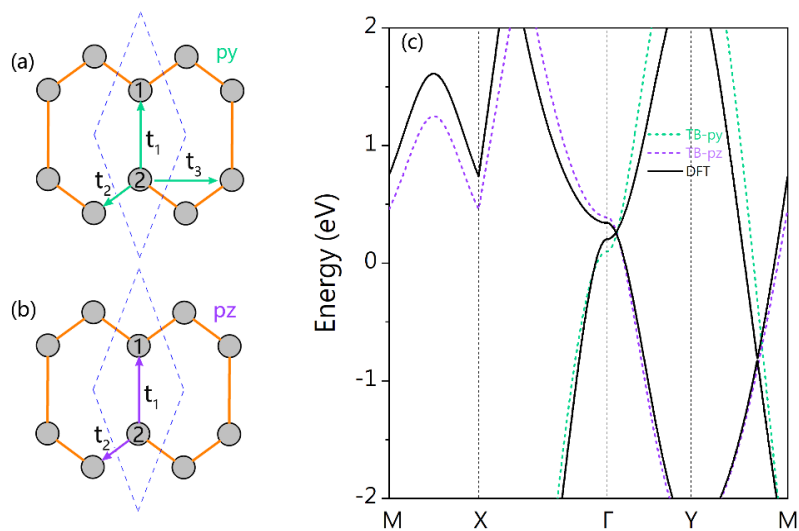

Figure 6. (a) and (b) indicate the hoppings considered for p$_y$ and p$_z$ orbitals in the TB model. The hopping parameters are labeled. (c) The comparison of TB model fitting and the first-principles

calculations. The green and purple dashed lines represent the contributions from $p_y$ and $p_z$ states respectively. The values of fitted hopping parameters are presented in Table 2.

Table 2. The tight-binding parameters for the TB model. The corresponding hopping processes are labeled in Fig. 6(a, b).

|       | $t_1$ | $t_2$ | $t_3$ | $\epsilon_\alpha$ |
|-------|-------|-------|-------|-------------------|
| $p_y$ | 1.5   | -1.9  | 0.7   | -3.7              |
| $p_z$ | 1.35  | -1.78 |       | 2.6               |

We construct a tight-binding (TB) model to describe the low-energy band structure of h-$B_2O$. Although there are two B atoms and one O atom in the primitive cell (see Fig. 1a), we find that the model may be simplified by dropping the O sites, which does not affect the symmetry of the model. Hence, we can write down a TB model based on the $p_y$ and $p_z$ orbitals on the B sites in the unit cell, as shown in Figure 6(a, b). The TB Hamiltonian reads:

$$H = \sum_{<i,j>} t_{i\alpha,j\alpha} c^\dagger_{i\alpha} c_{j\alpha} + \sum_i \epsilon_\alpha c^\dagger_{i\alpha} c_{i\alpha}, \quad (i,j \in \{1,2\}, \alpha \in \{p_y, p_z\}) \qquad (3)$$

where $c^\dagger_{i\alpha}$ and $c_{j\alpha}$ represent the creation and the annihilation operators, respectively. $\epsilon_\alpha$ is the on-site energy of the orbital $\alpha$; $t_{i\alpha,j\alpha}$ is the hopping energy between orbitals $\alpha$ at sites i and j. Since the $p_y$ and $p_z$ orbitals are orthogonal, it turns out that they are decoupled in the TB model (3). Only the nearest-neighbor hopping processes are considered for $p_z$ orbitals (see Fig. 6b). As for $p_y$, because the $p_y$ orbital lies in the plane of h-$B_2O$, the next-nearest-neighbor interaction cannot be ignored (see Fig. 6a). As shown in Fig. 6c, even with this very simple TB model, the main features of first-principles results can be captured very well.

As discussed before, h-$B_2O$ has good mechanical properties, hence lattice strain can be used as a convenient tool to tailor the electronic properties of h-$B_2O$. From the stress-strain relationship, we find that h-$B_2O$ can sustain at least 10% strain along the two lattice directions (see Fig. S7). The energy difference between $p_y$-band and $p_z$-band under different strains are plotted in Figure 7. With tensile strain along the a-axis, the nodal loop would contract towards the Y point (see Fig. 7a). On the other hand, with compressive strain along the a-axis, the loop expands towards the BZ center, and touch at the Γ point under a critical compressive strain of 1.45% (Fig. 7b). After passing through this critical strain, the loop splits into two loops, each penetrating through the BZ in the Γ–Y direction (see Fig. 7c). As discussed by Li *et al.*[52], a nodal loop circling around a point in the BZ is topologically distinct from a loop that traversing the BZ. The former can be contracted into a single point with preserved symmetry, whereas the latter cannot[52]. The distinction can be captured by the fundamental homotopy group for the BZ. In 2D, it is characterized by two integers $\pi_1(\mathbb{T}^2) = \mathbb{Z} \times \mathbb{Z}$, each integer indicating the number of times the loop winds around the BZ in a particular direction. In this sense, the loop in the equilibrium state of h-$B_2O$ is characterized by the topological numbers (0, 0). And the two loops in Fig. 7c are characterized by (0, ±1) (by specifying the winding direction of a loop with +1, its time-reversal partner would have -1). Thus, the strain-induced transformation may also be regarded as a topological phase transition. Strain applied along the b-axis can also induce topological phase transition. With appropriate strain, the same type of nodal-loop transformation is still revealed (see Fig. S8).

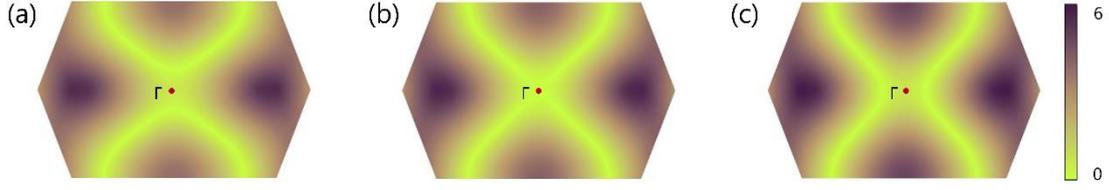

Figure 7. The energy difference of $p_y$-band and $p_z$-band for (a) 5% tensile strain, (b) 1.45% compressive strain, and (c) 5% compressive strain along the **a** direction. The glowing yellow lines denote the location of the nodal loops in the BZ.

Before closing, we would give a few remarks. First, compared with many examples of 3D nodal loop materials, 2D nodal loop materials are much less. Although there are quite a few theoretical proposals of 2D nodal line materials[54-60], only $Cu_2Si$ and CuSe have been confirmed in experiment[61, 62]. Still, $Cu_2Si$ and CuSe are not ideal, because the SOC in these materials are sizable and the loop is gapped by the SOC. In comparison, the nodal loop in h-$B_2O$ is more robust due to the negligible SOC. Moreover, the loop here is close to a critical point where a single loop can be split into two loops traversing the BZ. To our knowledge, such interesting feature has not been found before. (Although 2D loop traversing the BZ has been observed, e.g., in monolayer $X_3SiTe_6$ ($X$ = Ta, Nb)[58], the loop there is protected by the nonsymmorphic symmetry and hence pinned at the BZ boundary).

Second, a protected band crossing generally requires that the two bands belong to different irreducible representations (IRR). In the presence of sizable SOC, symmetry groups have to be extended to double groups[63]. Typically, the number of double valued IRRs is smaller than that of single valued ones. As a result, the number of IRRs actually is decreased when SOC is considered. This means a stable band crossing in the absence of SOC is likely to become unstable when SOC is included. This is exactly the reason of gap opening for $Cu_2Si$ and $CuSe$[61, 62]. From this reasoning, we may argue that light-element materials have a higher chance to host topological semimetal phases.

Finally, light-element materials also have another advantage: its band dispersion is typically larger. This is beneficial because the linear dispersion around the band crossing would expand a larger energy window, facilitating the experimental detection. We have calculated the band structure of h-$B_2O$ on substrate. As shown in Figure S11, the nodal loop feature in h-$B_2O$ can be maintained when encapsulated by 2D BN (see Supporting Information in detail).

## Conclusion

In conclusion, by using first-principles calculations, we discover a new 2D material, the hexagonal borophene oxide. The material is a completely flat monolayer, unlike the graphene or silicene oxides. We have confirmed the dynamic, thermal, and mechanical stability of h-$B_2O$. With the incorporated O atoms, h-$B_2O$ exhibits a remarkable mechanical anisotropy. Most importantly, h-$B_2O$ features a mirror symmetry protected nodal loop near the Fermi level, and there is no other extraneous band around. Interestingly, with moderate strain, the single loop can be transformed into two loops, each traversing the BZ, signaling a topological phase transition. We also point out that light-element materials have their own advantages for realizing topological semimetal phases. Our work not only reveals a new 2D material with interesting properties, it also provides an alternative route for achieving topological phases in two dimensions.

# Methods

Our first-principles calculations were performed within the density functional theory (DFT) formalism as implemented in the VASP code[64, 65]. We used the generalized gradient approximation with the Perdew-Burke-Ernzerhof (PBE)[66] realization. The interaction between the core and valence electrons was described by the projector-augmented wave method[67]. The kinetic energy cutoff was set to 600 eV. The atomic positions were optimized using the conjugate gradient method with total energy and maximum force convergence criteria set as $10^{-6}$ eV/atom and $10^{-3}$ eV/Å, respectively. A $2\pi \times 0.02$ Å$^{-1}$ k-point density grid according to Monkhorst-Pack scheme was used to sample the BZ. The vacuum region between adjacent images in the direction normal to the monolayer plane was set at about 15 Å. Moreover, hybrid functional calculations of HSE06[68] was used to confirm the electronic band structures. In the calculation of phonon spectra, we employed the finite difference method as implemented in the Phonopy[69] code with the forces calculated from VASP. AIMD simulation under the canonical (NVT) ensemble was performed. We visualized the atomic structures with VESTA package[70].

# Acknowledgements

The authors thank Pan Zhou for his valuable discussion. This work is supported by National Natural Science Foundation of China (Grant Nos. 11804039, 11804040, 11804041), and by the Singapore Ministry of Education AcRF Tier 2 (Grant No. MOE2015-T2-2-144).